\documentclass[aps,prl,showpacs,floatfix,twocolumn,superscriptaddress]{revtex4-1}
\usepackage{graphicx}
\usepackage{bm,amsmath}
\usepackage{natbib}
\usepackage{dcolumn}
\usepackage{subfigure}
\usepackage{multirow}
\usepackage{mathrsfs}
\begin{document}

\title{Anomalous quantum-critical scaling corrections in two-dimensional antiferromagnets}

\author{Nvsen Ma}
\affiliation{State Key Laboratory of Optoelectronic Materials and Technologies, School of Physics, 
Sun Yat-Sen University, Guangzhou 510275, China}
\affiliation{Beijing National Laboratory of Condensed Matter Physics and Institute of Physics, Chinese Academy of Sciences, Beijing 100190, China}
\affiliation{Department of Physics, Boston University, Boston, Massachusetts 02215, USA}

\author{Phillip Weinberg}
\affiliation{Department of Physics, Boston University, Boston, Massachusetts 02215, USA}

\author{Hui Shao}
\affiliation{Beijing Computational Science Research Center, Beijing 100193, China}
\affiliation{Department of Physics, Boston University, Boston, Massachusetts 02215, USA}

\author{Wenan Guo}
\affiliation{Department of Physics, Beijing Normal University, Beijing 100084, China}
\affiliation{Beijing Computational Science Research Center, Beijing 100193, China}

\author{Dao-Xin Yao}
\email{yaodaox@mail.sysu.edu.cn}
\affiliation{State Key Laboratory of Optoelectronic Materials and Technologies, School of Physics, 
Sun Yat-Sen University, Guangzhou 510275, China}

\author{Anders W. Sandvik}
\email{sandvik@bu.edu}
\affiliation{Department of Physics, Boston University, Boston, Massachusetts 02215, USA}
\affiliation{Beijing National Laboratory of Condensed Matter Physics and Institute of Physics, Chinese Academy of Sciences, Beijing 100190, China}

\date{\today}

\begin{abstract}
We study the N\'eel--paramagnetic quantum phase transition in two-dimensional dimerized $S=1/2$ Heisenberg antiferromagnets using 
finite-size scaling of quantum Monte Carlo data. We resolve the long standing issue of the role of cubic interactions arising in the 
bond-operator representation when the dimer pattern lacks a certain symmetry. We find non-monotonic (monotonic) size dependence in 
the staggered (columnar) dimerized model, where cubic interactions are (are not) present. We conclude that there is a new irrelevant 
field in the staggered model, but, at variance with previous claims, it is not the leading irrelevant field. The new exponent is 
$\omega_2 \approx 1.25$ and the prefactor of the correction $L^{-\omega_2}$ is large and comes with a different sign from that of the 
conventional correction with $\omega_1 \approx 0.78$. Our study highlights competing scaling corrections at quantum critical points.
\end{abstract}

\maketitle

One of the best understood quantum phase transitions is that between N\'eel antiferromagnetic (AFM) and quantum paramagnetic ground 
states in bipartite two- and three-dimensional (2D and 3D) dimerized Heisenberg models with inter- and intra-dimer couplings $J_1$ and $J_2$
\cite{sigma,3dmap,millis,chubukov94,always,thebook}. The ground state hosts AFM order when $g=J_2/J_1 \approx 1$, and there is a 
critical point at some model dependent $g_c >1$. The 3D version of this transition for $S=1/2$ spins has an experimental realization 
in TlCuCl$_3$ under high pressure \cite{merchant14,qin15}. While no 2D realization exists as of yet (though the magnetic field driven 
transition has been realized \cite{bec}), this case has been very important for developing 
the framework for 2D quantum phase transitions of the N\'eel AFM state \cite{gc}. The field theory of the AFM--paramagnetic transition 
is now well developed, and efficient quantum Monte Carlo (QMC) methods can be used to study ground states of microscopic models with tens of 
thousands of spins \cite{thebook}. Many non-trivial predictions for scaling in temperature, frequency, system size, etc., have been tested 
\cite{sandvik94,troyer1996,ladder,bilayer,sen15,lohofer15}.

\begin{figure}[b]
\begin{center}
\includegraphics[width=68mm, clip]{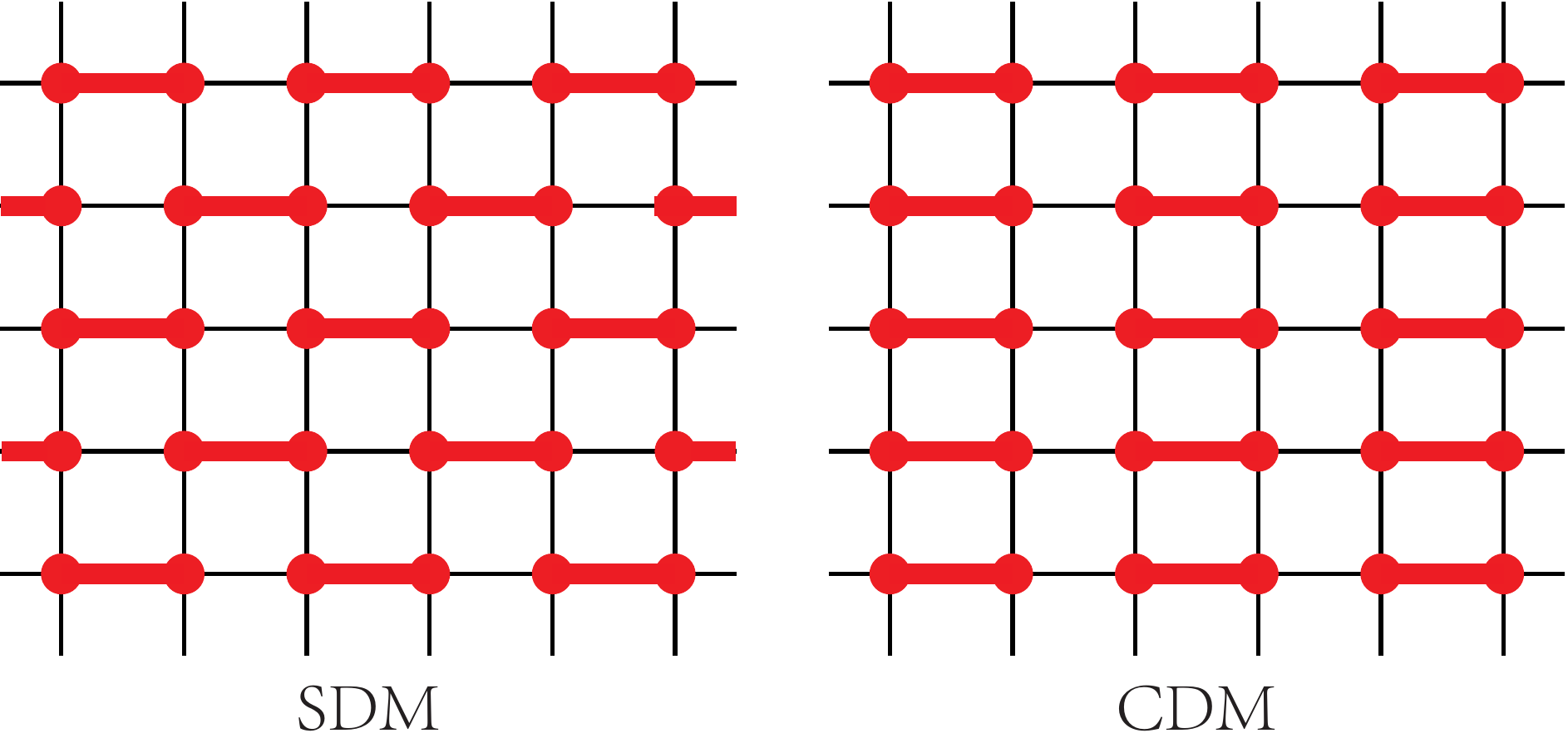}
\end{center}
\vskip-5mm
\caption{The Heisenberg SDM and CDM studied in this work. Black (thinner) and red (thicker) bonds represent intra- and inter-dimer 
exchange ${\bf S}_i \cdot {\bf S}_j$, of strength (prefactor) $J_1$ and $J_2$, respectively, between $S=1/2$ spins.}
\label{model}
\end{figure}

Despite many successes, there are still questions surrounding the 2D AFM--paramagnetic transition. A long-standing unresolved issue is 
differences observed in QMC calculations between two classes of dimer patterns \cite{different,jiang09,unusual,cubic,accurate}, 
exemplified by the often studied columnar dimer model (CDM) and the initially less studied staggered dimer model (SDM), both illustrated in 
Fig.~\ref{model}. Indications from finite-size scaling of a universality class different from the expected 3D O(3) class in the SDM \cite{different} 
led to several follow-up studies \cite{jiang09,unusual,cubic,accurate}. The consensus now is that there is no new universality class, as defined by the 
standard critical exponents. However, because of the lack of a certain local symmetry, cubic interactions arise in 
the bond-operator description of the SDM, which in the renormalization group corresponds to an irrelevant field that is present neither in the 
CDM nor in the classical O(3) model \cite{cubic}. Thus, the SDM contains an interesting quantum effect worthy of further investigations. 

In this Letter we report detailed comparisons of the finite size ($L$) scaling corrections of type $L^{-\omega}$ in the CDM and SDM. While 
previous works on judiciously chosen observables \cite{unusual} and lattices with optimized aspect ratios \cite{accurate} have convincingly 
demonstrated O(3) universality, the reasons for the unusual scaling behaviors of the SDM have never been adequately 
explained. In Ref.~\onlinecite{cubic}, QMC calculations indicated that the exponent of the leading correction is smaller than in the CDM, but 
the value, $\omega \approx 0.6$ in the SDM \cite{cubic,accurate} versus the conventional value $\omega \approx 0.78$ \cite{guida98,besteta} 
in the O(3) model and the CDM, is not very different and cannot explain all the observed anomalous finite-size scaling 
properties of the SDM.

We here study $L \times L$ CDM and SDM systems of size up to $L=256$. Focusing on the scaling corrections, we fix the leading critical exponents at 
their known O(3) values in our finite-size analysis, which allows us to reliably investigate also subleading corrections. In contrast to the 
previous studies, we demonstrate that the SDM actually does not have a smaller $\omega_1$ 
than the CDM. Instead, the cubic interaction induces the next correction, which has $\omega_2 = 1.25(3)$ (where the number 
within parathesis here and henceforth denotes the statistical error in the preceding digit) and a large prefactor of sign different
from that of the first correction. This causes non-monotonic finite-size behaviors that were previously either not observed \cite{unusual,cubic} 
or not analyzed properly \cite{accurate}. 

{\it QMC and fitting procedures.}---We here use the standard stochastic series expansion QMC method \cite{sandvik99,thebook} for $S=1/2$ spins and 
set the inverse temperature $\beta$ at $L/2$ (so that $L/\beta$ is close to the spinwave velocity \cite{accurate}). At a quantum phase transition 
with dynamic exponent $z=1$ (as is the case here), as long as $\beta \propto L$ the temperature does not appear as an independent argument in the 
scaling function obtained from renormalization group theory. In the case of a dimensionless quantity we have \cite{barber,compost}
\begin{equation}
O(g,L)=f[(g-g_c)L^{1/\nu},\lambda_{1}L^{-\omega_{1}},\lambda_{2}L^{-\omega_{2}}, \cdots],
\label{first}
\end{equation}
if $g$ is sufficiently close to $g_c$. Here $\lambda_i$ denotes the irrelevant fields, which we order such that $\omega_{i+1} > \omega_i > 0$.
Useful dimensionless quantities to study in QMC calculations include the Binder ratio $R=\langle m_{z}^{4}\rangle/{\langle m_{z}^{2} \rangle^{2}}$,
where $m_{z}$ is the component of the staggered magnetization along the quantization axis, the $L$-normalized spin stiffness constants 
$L\rho_x$ and $L\rho_y$ (with $x$ and $y$ referring to the lattice directions), and the uniform susceptibility $L\chi_u$. 
We refer to Ref.~\onlinecite{thebook} for technical details. 

To linear order in the first irrelevant field, Eq.~(\ref{first}) can be written as
\begin{equation}
O(g,L)=f_0(\delta L^{1/\nu})+L^{-\omega_{1}}f_1(\delta L^{1/\nu}),
\label{leading}
\end{equation}
where $\delta=g-g_c$ and $f_0$ and $f_1$ are scaling functions related to the original $f$. Thus, in the absence of corrections ($f_1=0$), 
a dimensionless quantity is size independent at $g_c$, and by expanding $f_0$ we see that $O(g,L)$ for different $L$ cross each
other at $g_c$. With the scaling correction included, the crossing points only drift toward $g_c$ as $L \to \infty$, and for two different
sizes $L$ and $L'=rL$ one can derive simple expressions for the crossing value $g^*(L)$ and the observable
$O^*(L)$ at this point \cite{luck};
\begin{subequations}
\begin{align}
&g^{*}(L) = g_c + aL^{-\omega_{1}-1/\nu}, \label{crossa} \\
&O^*(L) = O_c + bL^{-\omega_{1}} \label{crossb},
\end{align}
\label{cross}
\end{subequations}
where only $a$ and $b$ depend on $r$. We use $r=2$ as a convenient size 
ratio allowing for a large number of size pairs $(L,2L)$, with size series of the form $L=s2^n$ for a range of integers 
$n$ and several choises of $s$. Tests with other $r$ reveal no changes in the asymptotics.

\begin{figure}
\begin{center}
\includegraphics[width=75mm]{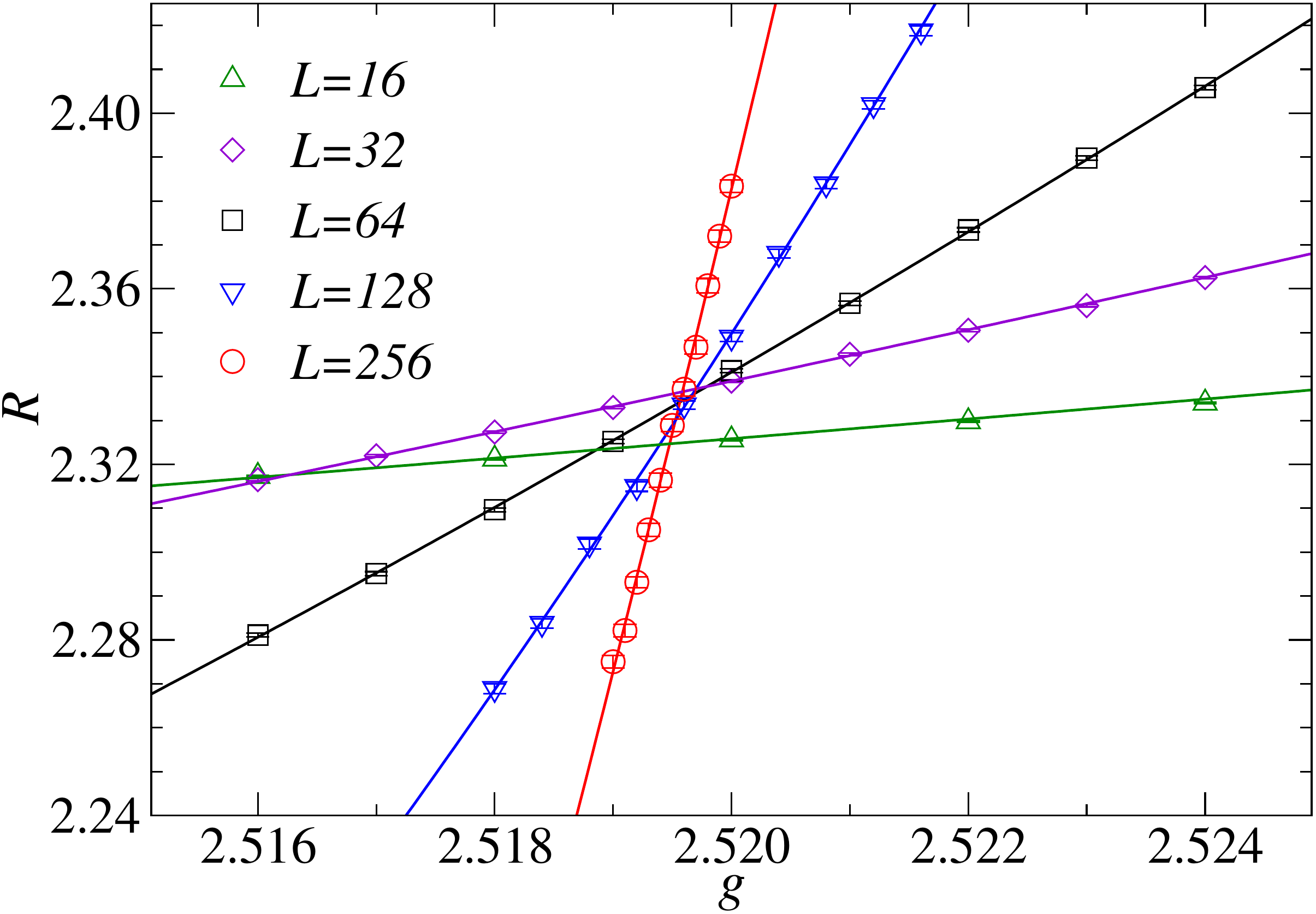}
\end{center}
\vskip-3mm
\caption{Binder ratio of the SDM for several system sizes in the neighborhood of $g_c$. The curves are polynomial fits
giving crossing points $(g^*,R^*)$ between $(L,2L)$ data.}
\label{r2cross}
\vskip-2mm
\end{figure}

We extract the crossings using third-order polynomial fits to ten or more data points in the neighborhood 
of $g_c=g^*(\infty)$, with the window $[g_{\rm min},g_{\rm max}]$ reduced as $L$ is increased. Such interpolations give reliable 
crossing points, and statistical errors are computed using bootstrapping. Examples of data with fits are shown in 
Fig.~\ref{r2cross}.

When fitting the crossing points $g^*(L)$ and $O^*(L)$ to their appropriate finite-size scaling forms, 
the same system size $L$ can appear in two pairs, $(L,2L)$ as well 
as $(L/2,L)$. There are therefore some covariance effects, which we take into account by using the full covariance matrix (computed using bootstrap 
analysis) in the definition of the goodness of the fit $\chi^2$. When jointly fitting to two different
quantities, we also account for the associated covariance. For the functional forms, we will go beyond the first-order expansion 
leading to Eqs.~(\ref{cross}), and this will be the key to our findings and conclusions. 

\begin{figure}[t]
\begin{center}
\includegraphics[width=84mm]{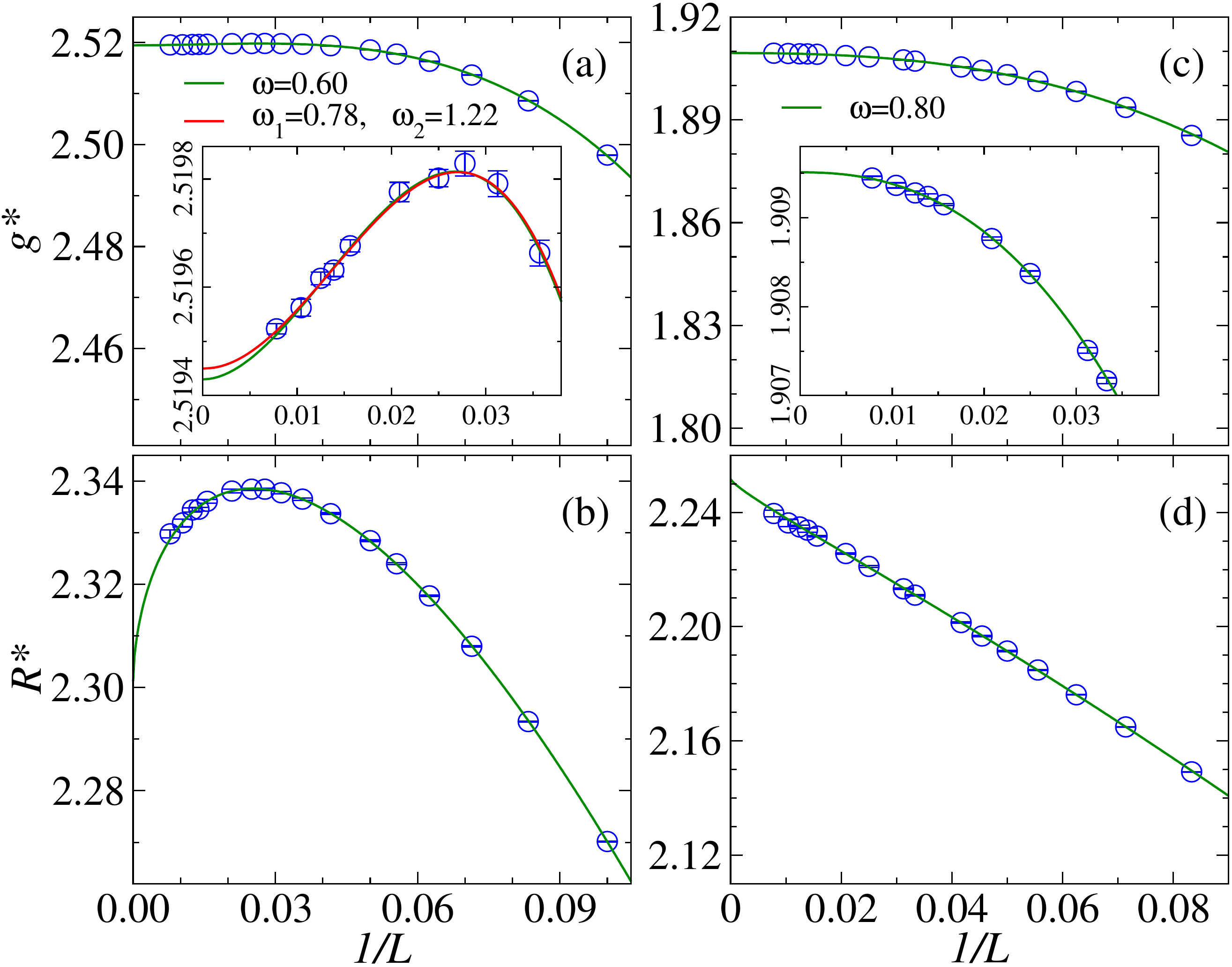}
\caption{Inverse system size dependence of $(L,2L)$ crossing data for the SDM (a,b) and the CDM (c,d) along with joint fits (green curves)
of the forms in Eq.~(\ref{joint}). The exponent $\omega$ is adjusted for optimal fits, giving $\omega=0.60(4)$ for the SDM and $\omega=0.80(2)$ for 
the CDM. The insets show the large system data on more detailed scales. The red curve in the inset of (a) shows a fit with only the leading terms
arising from the first and second irrelevant fields, with $\omega_1=0.78$ fixed and $\omega_2=1.22(5)$ resulting from the fit; the corresponding
fitting curve in (b) barely changes and is not shown.}
\label{gcandr2}
\end{center}
\end{figure}

{\it Finite-size scaling.}---The size dependence of $R$ crossing points is shown in Fig.~\ref{gcandr2} for both models. A striking feature 
is the non-monotonic behaviors apparent for the SDM but not present for the CDM. Note here that $1/L$ on the horizontal axis refers to the smaller 
of the two system sizes $(L,2L)$ used for the crossing points, and the maximums in $g^*$ and $R^*$ are at $2L \approx 80$. In the original discovery of the anomalous 
behaviors for the SDM \cite{different}, the systems were smaller and the correct asymptotic behaviors were therefore not reached.

We will first assume that only one irrelevant field is important but treat the corrections beyond the first-order expansion in $L^{-\omega_1}$, 
Eq.~(\ref{leading}). Later we will argue that one has to include also the $L^{-\omega_2}$ term in the case of the SDM, while for the CDM $\omega_2$ 
is much larger and does not have to be considered. Even with only one irrelevant field, if the associated exponent $\omega=\omega_1$ is 
small, the higher order terms such as $L^{-2\omega}$ will also be important. As a guide to how far to go, we here compare the previous
estimates $\omega_1 \approx 0.5 - 0.6$ \cite{cubic,accurate} in the SDM with the second correction of the O(3) model, with $\omega_2 \approx 1.8$ 
\cite{newman84}, and note that several additional corrections with exponents close to $2$ are expected \cite{hasen01}. It would then be 
pointless to go to higher order than $3\omega$ in the first irrelevant field, and with $1/\nu \approx 1.4$ we also do not include mixed 
corrections with $\omega$ and $1/\nu$. Thus, for the SDM we use
\begin{subequations}
\label{joint}
\begin{align}
g^*(L)=&g_{c}+L^{-{1/\nu}}(a_{1}L^{-\omega}+a_{2}L^{-2\omega} + a_{3}L^{-3\omega}), \label{joint_g} \\
R^*(L)=&R_c+b_{1}L^{-\omega}+b_{2}L^{-2\omega} +b_{3}L^{-3\omega}, \label{joint_r}
\end{align}
\end{subequations}
and exclude small systems until good fits are obtained. For the CDM, with $\omega_1 = 0.78$, by the above arguments we stop at $2\omega$.

The fitting coefficients $a_{i}$ and $b_{i}$ in Eq.~(\ref{joint}) are not fully independent of each other but are related because they 
originate from the same scaling function, Eq.~(\ref{first}). We do not write down the relationships here but fully take them into 
account in joint fits of the $g^*$ and $O^*$ data. These nonlinear fits are quite demanding and we make use of a slow but reliable stochastic approach 
\cite{optim}. The stability of the fits is greatly aided by fixing $1/\nu$ to its known 3D O(3) value $1.406$ \cite{besteta}. 
The resulting curves are shown in Fig.~\ref{gcandr2}. Here, as in all cases below, all data points shown in the figure were included in the fits 
(with smaller sizes excluded until the fits have acceptable $\chi^2$ values).

For the CDM, our result for the critical coupling is $g_{c}=1.90951(1)$. The value is consistent with the best previous results, $g_{c}=1.90948(4)$ 
\cite{thebook} and $g_{c}=1.90947(3)$ \cite{accurate}, but with reduced statistical error. For the correction, we obtain $\omega=0.80(2)$, which 
agrees with the O(3) value $\omega_1=0.782(13)$ \cite{besteta}. 

For the SDM we obtain $g_{c}=2.51943(1)$. Using rectangular lattices with optimized aspect ratio, a compatible result, $g_c=2.51941(2)$, was obtained 
\cite{accurate}. For the correction we obtain $\omega=0.60(4)$, which is clearly smaller than the known O(3) value cited 
above but in good agreement with the values presented in both Refs.~\cite{cubic} and \cite{accurate}. 

\begin{table}[b]
\caption{Results for the critical point and correction exponent obtained from fits of various dimensionless quantities to scaling 
forms analogous to Eqs.~(\ref{joint}), keeping corrections up to $3\omega$ for the SDM and $2\omega$ for the CDM.}
\vskip1mm
\begin{tabular}{|c|c|c|c|c|}
\hline
\multirow{2}{*}{}&\multicolumn{2}{|c|}{SDM} & \multicolumn{2}{|c|}{CDM}\\
\hline
&$\omega$&$g_{c}$&$\omega$&$g_{c}$\\
\hline
~$L\rho_{x}$& $0.88(2)$ & $2.51946(2)$&$0.77(3)$& $1.90953(2)$\\
\hline
$L\rho_{y}$& $0.39(5)$ & $2.51942(3)$ & $0.77(4)$ & $1.90957(2)$ \\
\hline
$L\chi_{u}$&$ 0.68(6)$ & $2.51945(2)$ &$ 0.78(3)$ & $1.90956(3)$ \\
\hline
$R$ & $0.60(4)$ & $2.51943(1)$ &$ 0.80(2)$ & $1.90951(1)$\\
\hline
\end{tabular}
\label{allcri}
\end{table}

Although $R_c$ is universal in the sense that it does not depend on the micro structure of lattice and details of the interactions, it does
depend on boundary conditions \cite{kamienartz,selke} and aspect ratios \cite{accurate}. The CDM and SDM have different critical spinwave velocities 
and, therefore, effectively different time-space aspect ratios even though $\beta/L$ is the same. This explains the different $R_c$ values
in Fig.~\ref{gcandr2}; see also Supplemental Material \cite{smcite}.

By analyzing also the spin stiffness and the uniform susceptibility in the manner described above, we obtain the results
summarized in Tab.~\ref{allcri}. The results for the CDM consistently reproduce the known O(3) value of $\omega_1$, 
while in the case of the SDM the different quantities produce a wide range of results. The latter suggests that $\omega$ 
may not be the true smallest correction exponent in the case of the SDM, but, as also pointed out in Ref.~\cite{cubic}, should be regarded as an 
``effective exponent'' influenced by neglected further corrections. The inability of a single irrelevant field to describe the
data is actually not unexpected within the scenario of irrelevant cubic interactions \cite{cubic}, because the standard leading correction 
with $\omega_1 \approx 0.78$ should still be present and may produce various ``effective'' scaling behaviors over a limited range of system 
sizes when combined with the cubic perturbation. Thus, a reliable analysis of the SDM should require at least 
$\omega_1$ and $\omega_2$.

\begin{figure}
\begin{center}
\includegraphics[width=84mm]{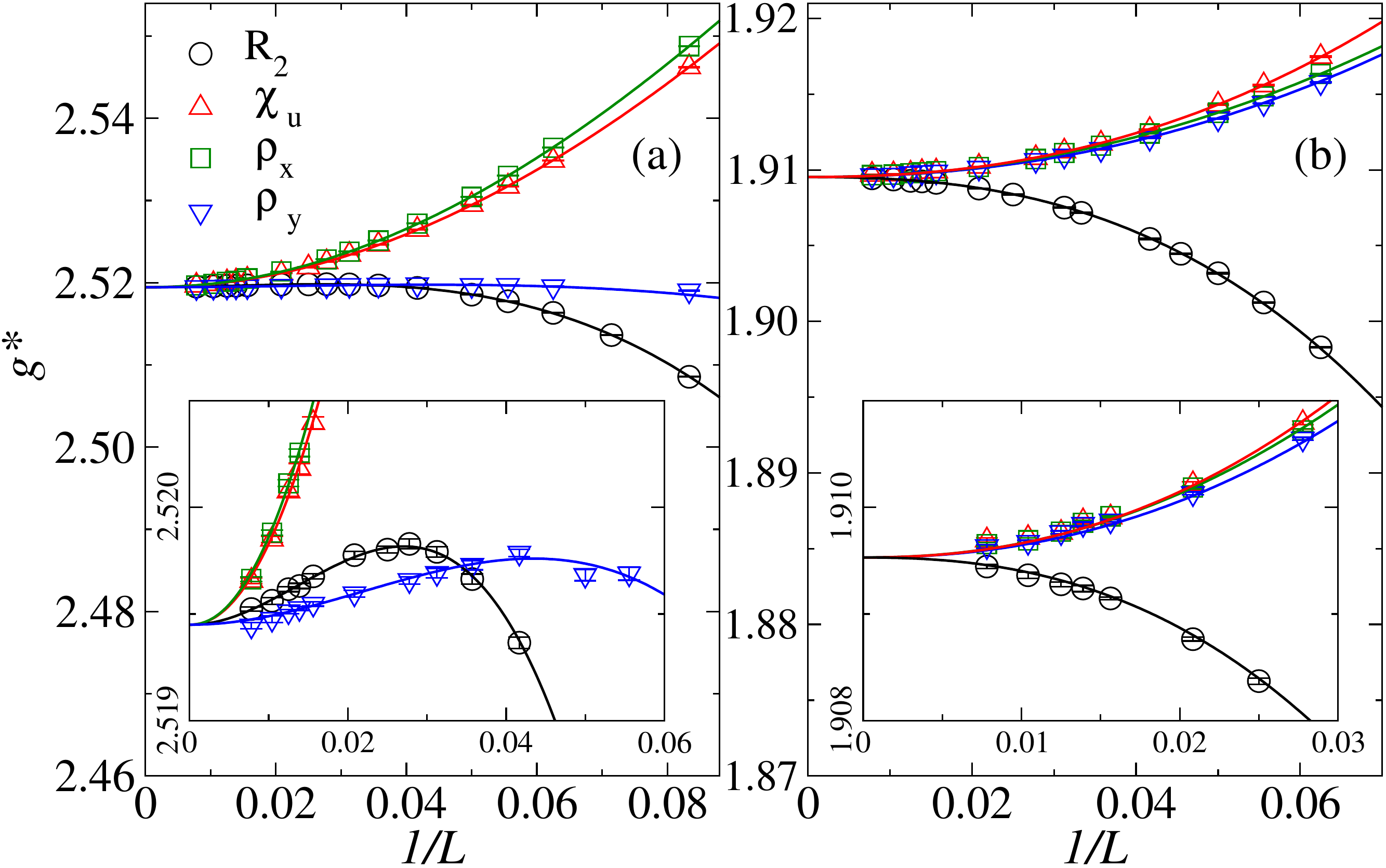}
\caption{Joint fits of crossing data for several quantities where $g^*(\infty)=g_c$ is fixed to a common value and two
corrections are used to first order, with  $\omega_{1}=0.78$ and $1/\nu=1.406$. The insets zoom in on the data for the larger 
system sizes. For the SDM (a), the fit delivers 
$g_{c}(\infty)=2.51945(1)$ and $\omega_{2}=1.30(7)$, $1.3(1)$, $1.2(1)$ and $1.0(2)$ from $R$, $L\chi_{u}$, $L\rho_{x}$, 
and $L\rho_{y}$, correspondingly. In the CDM fits (b), $2\omega_1=1.56$ was used in place of $\omega_2$ and $g_c=1.90956(2)$.}
\label{fixedgc}
\end{center}
\end{figure}

We can generalize Eqs.~(\ref{joint}) to two correction exponents, $\omega_1$ and $\omega_2$, but in that case it is difficult to 
determine both of them with sufficient precision. However, since the standard leading correction should still be present \cite{cubic}, 
we now also fix $\omega_1=0.78$ and only treat $\omega_2$ as a free parameter. It is then sufficient to go to linear
order in the corrections and yet obtain fully acceptable fits. We obtain  $g_{c}=2.51945(1)$ and $\omega_{2}=1.22(5)$ 
for the SDM. The new fitted curve is shown in the inset of Fig.~\ref{gcandr2}(a). The estimate of $g_c$ is a bit higher than 
the previous value from $R^*$, but the difference is not statistically significant.

The key result here is clearly that $\omega_2$ comes out larger than the leading O(3) exponent. It is, however, significantly smaller than the 
expected second irrelevant $O(3)$ exponent with value $\approx 1.8$ \cite{newman84,hasen01}, and it is also less than $2\omega_1$. The new correction 
should therefore be due to the cubic interactions \cite{cubic} in the low-energy theory of the SDM. To test the stability of $\omega_2$ across 
different quantities, we also used a slightly different procedure of fitting only to $g^*$ (instead of the joint fit with $R^*$) and requiring the 
same $L \to \infty$ value of $g_c$ for all the quantities considered. We still also fix $1/\nu=1.406$ and  $\omega_1=0.78$ but keep $\omega_2$ free 
for all individual quantities. The SDM data with fits are displayed in Fig.~\ref{fixedgc}(a), with the resulting $g_c$ and $\omega_2$ estimates 
listed in the caption. The fits are statistically good and all four $\omega_2$ estimates are consistent with the value obtained above.
In the case of the CDM, shown Fig.~\ref{fixedgc}(b), we follow the same procedures but replace $\omega_2$ by $2\omega_1$ and there is no free exponent. 
This fit is only of marginally acceptable statistical quality even when starting the fits from $L=16$, indicating some effects still of the higher-order terms 
that were included in Fig.~\ref{gcandr2}(b). We therefore keep the value from $R$ in Tab.~\ref{allcri} as our best $g_c$ estimate for this model.

\begin{figure}[t]
\begin{center}
\includegraphics[width=70mm]{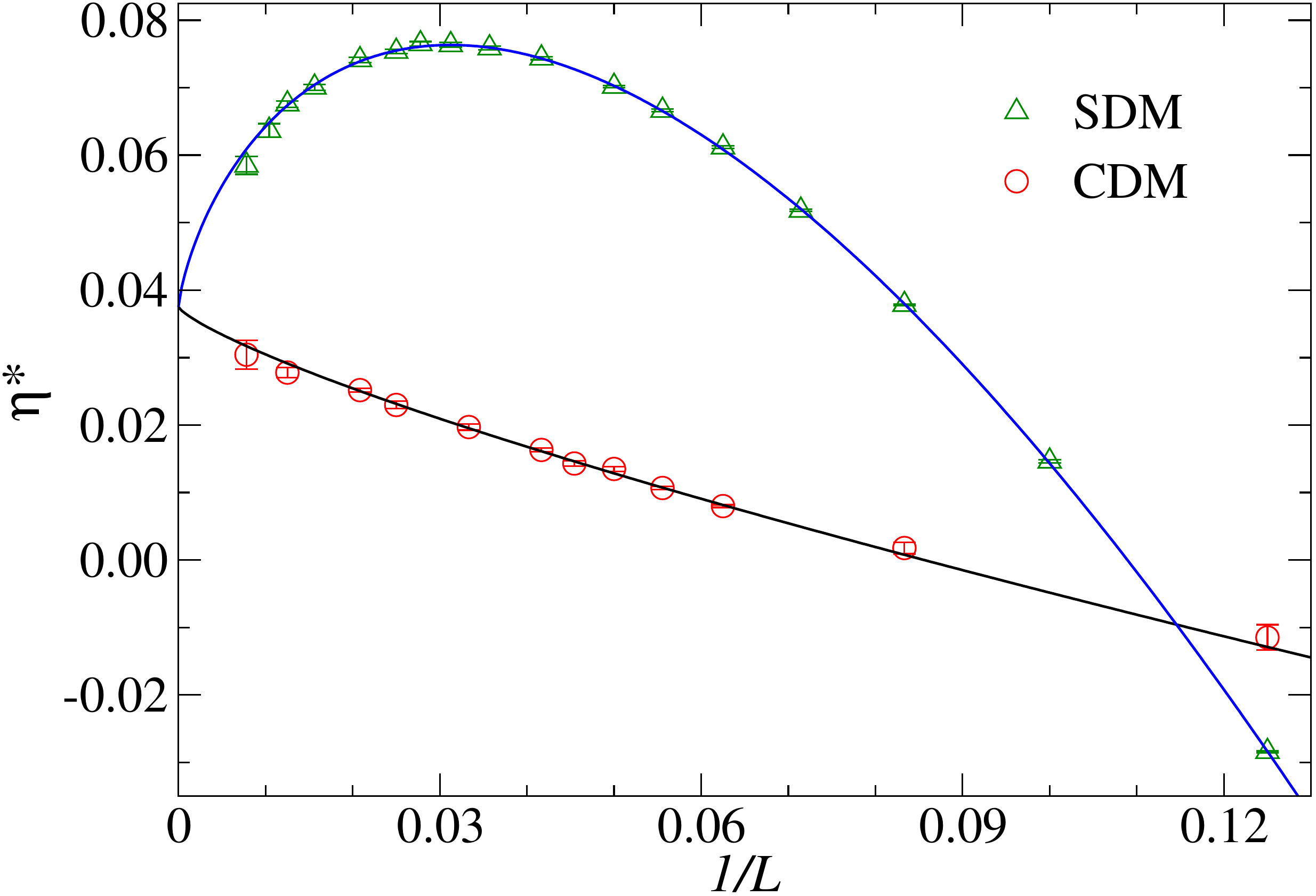}
\caption{Size dependence of the exponent $\eta$ as defined in Eq.~(\ref{caleta}). The known infinite-size value $\eta=0.0375$ is fixed
in the fits (curves). The CDM data are fitted with only the first correction term in Eq.~(\ref{caleta2}), with $\omega_1=0.78$
fixed. In the SDM fit $\omega_{1}=0.78$ is also fixed and the second exponent $\omega_{2}=1.29(5)$ is the result of the fit.}
\label{etaplot}
\end{center}
\end{figure}

To further ascertain our conclusions about the SDM, we also consider the squared order parameter itself. Having determined a precise 
estimate of $g_c$, we study the scaling of $\langle m^2\rangle$ at this point, where we expect 
\begin{equation}
\langle m^{2}\rangle_c \propto L^{-(1+\eta)}(1+b_{1}L^{-\omega_{1}} + b_{2}L^{-\omega_{2}} + \ldots).
\label{eta2}
\end{equation}
We can then define a size-dependent exponent as
\begin{equation}
\eta^*(L)=\ln[\langle m^{2}(L)\rangle_c/\langle m^{2}(2L)\rangle_c]/\ln(2) -1,
\label{caleta}
\end{equation}
which should scale as
\begin{equation}
\eta^*(L)=\eta+c_{1}L^{-\omega_{1}} +c_{2}L^{-\omega_{2}} + \ldots.
\label{caleta2}
\end{equation}
To test this form and extract $\omega_2$, we use the known value $\eta=0.0375(5)$ \cite{besteta} and fix $\omega_1=0.78$. As shown 
in Fig.~\ref{etaplot}, the form fits the data very well and gives $\omega_2=1.29(5)$. Here one can again see how access to only system 
sizes less than $L=80$ could easily lead to the wrong conclusion. A fit with two adjustable exponents gives 
$\omega_1 = 0.77(6)$ and $\omega_2=1.31(7)$, perfectly consistent with the fit with $\omega_1$ fixed. In the case of the CDM, 
also shown in Fig.~\ref{etaplot}, we find that the data are well described with a single correction with the known value of the exponent.

{\it Conclusions.}---We have analyzed the SDM under the scenario \cite{cubic} of an O(3) quantum phase transition with an additional 
irrelevant perturbation that is absent in the CDM. Our results are consistent with this picture and demand a new scaling correction with 
exponent $\omega_2 \approx 1.25$ that is larger than the also present conventional 3D O(3) exponent $\omega_1 \approx 0.78$ but smaller than 
the next known O(3) exponent. Thus, the cubic interactions in the low-energy theory are formally more irrelevant than previously believed
\cite{cubic,accurate}, but their effects are important in finite-size scaling of many quantities because of their different signs and 
larger prefactors of the correction terms (four times larger than the factor of the leading correction in the case of the order
parameter), thus giving rise to  non-monotonic behaviors. 

In addition to resolving the role of the cubic interactions
in the class of models represented by the SDM, our study also serves as an example of finite-size behaviors that may at first sight appear
puzzling but can be understood once the possibility of competing scaling corrections is recognized. Nonmonotonic scaling has also been observed 
at the deconfined quantum phase transitions, which has complicated efforts to extract the critical point and exponents \cite{shao16}.

\begin{acknowledgments}
{\it Acknowledgments.}---We would like to thank Ning Su, Stefan Wessel, and Matthias Vojta for useful discussions.
The work of N.M. and D.X.Y. was supported by Grants No.~NKRDPC-2017YFA0206203, No.~NKRDPC-2018YFA0306001,
No.~NSFC-11574404, No.~NSFC-11275279, No.~NSFG-2015A030313176, and the Leading Talent Program of Guangdong Special Projects. 
H.S. was supported by the China Postdoctoral Science Foundation under Grants No.~2016M600034
and No.~2017T100031. W.G. was supported by NSFC under Grants No.~11775021 and No.~11734002. A.W.S was supported by the NSF under Grant
No.~DMR-1710170 and by a Simons Investigator Award. Some of the calculations were carried out on Boston University's Shared Computing Cluster.
\end{acknowledgments}

\newpage

\setcounter{page}{1}
\setcounter{equation}{0}
\setcounter{figure}{0}
\renewcommand{\theequation}{S\arabic{equation}}
\renewcommand{\thefigure}{S\arabic{figure}}

\section{Supplemental Material}

\begin{center}
{\bf \noindent
Anomalous quantum-critical scaling corrections in two-dimensional antiferromagnets}
\vskip2mm

{\noindent
N. Ma, P. Weinberg, H. Shao, W. Guo, D.-X. Yao, and A. W. Sandvik}
\vskip2mm
\end{center}
\vskip2mm

Here we discuss the dependence of the critical Binder ratio $R_c$ on the time-space aspect ratio $\beta/L$ of the system in the QMC simulations, 
to explain the fact that the results for the SDM and the CDM in Fig.~3 of the main text do not extrapolate to the same value when 
$L \to \infty$. We also comment more broadly on the role of aspect ratios when analyzing quantum phase transitions.

The dependence of the Binder ratio on the spatial aspect ratio in classical systems is well understood \cite{kamienartz,selke}, and
in quantum systems $\beta/L$ also acts as an aspect ratio. In addition, the CDM and SDM lack 90$^\circ$ lattice rotational invariance
and therefore have different velocities of excitations in the two lattice directions. In order to obtain the universal value of
$R_c$, one has to find both the correct spatial aspect ratio $L_y/L_x$, corresponding to the ratio of the two velocities, and the 
temporal ratio $\beta/L$. This was done in Ref.~\cite{accurate}, and the $R$ crossing values of the CDM and SDM were shown to indeed 
be universal, agreeing with the value obtained for the 3D classical Heisenberg model at its critical temperature.

\begin{figure}[b]
\begin{center}
\includegraphics[width=70mm]{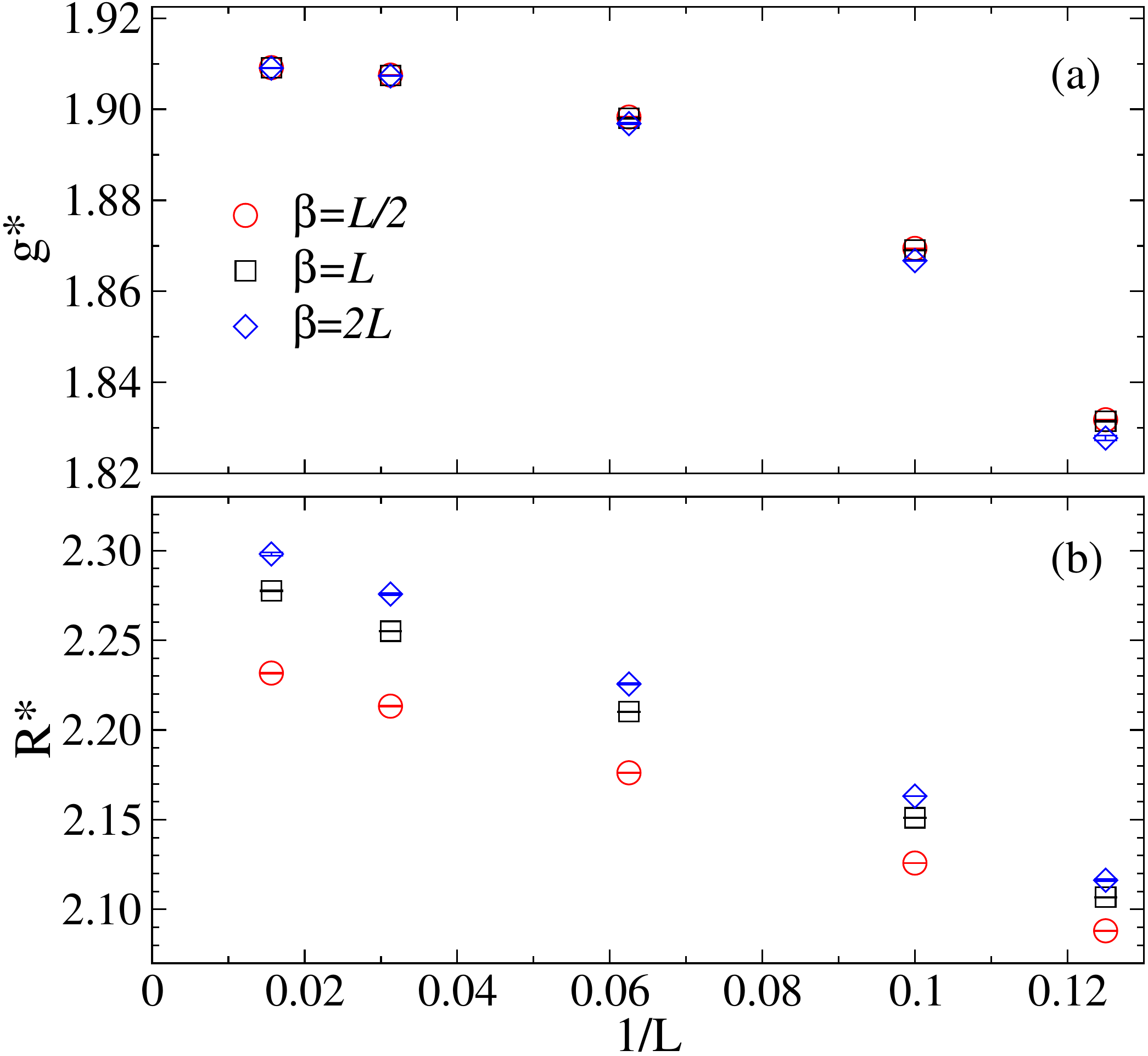}
\caption{Inverse size dependence of the Binder crossing points, (a) for the coupling ratio and (b) for the corresponding
value of the Binder ratio, obtained from system sizes $L$ and $2L$ for the CDM at different values of $\beta/L$.}
\label{betal}
\end{center}
\end{figure}

Here we just illustrate the dependence on the temporal ratio in the case of the CDM, keeping the $L \times L$ spatial geometry.
The results shown in Figure \ref{betal} demonstrate that the critical point consistently flows to the same value, while the asymptotic
$R$ crossing value depends on $\beta/L$. We do not extrapolate these results to infinite size, as the purpose is just to illustrate
the very clear flows toward incompatible infinite-size values for different $\beta/L$ ratios. Although the CDM and SDM have the same
$\beta/L$ ratio in the QMC simulations leading to the results in Fig.~\ref{gcandr2}, the effective aspect ratio is still different because
of the different spinwave velocities. The two models also have effectively different spatial aspect ratios.

While we agree with Ref.~\cite{accurate} on the point of the common universality of the CDM, SDM, and O(3) models, and the
importance of tuning aspect ratios if one desires to observe the universal Binder cumulant, we are not convinced of the practical
utility of finding the special aspect ratios and make the system effectively perfectly space-time isotropic. Optimizing the aspect 
ratios is an additional complication in the simulations, though potentially the symmetry between the directions also could have 
advantageous effects on the scaling, thoough this is not clear from the results presented so far. In Ref.~\cite{accurate} some 
non-monotonic behaviors were also seen, i.e., the corrections arising from the cubic interactions do not vanish at the special 
aspect ratios, which one should also not expect. As we have shown in the main text, one can reach the correct conclusions on
the universality class also with $L \times L$ lattices and with any fixed reasonable ratio $\beta/L$ (where one should also keep 
in mind that the QMC simulation time scales linearly with $L$ and with $\beta/L$). The key to understand fully the role of the cubic 
interactions in the SDM is to realize the importance of two irrelevant fields in the finite-size analysis.

\end{document}